\documentclass[pre,showpacs,preprint,superscriptaddress]{revtex4-1}
\usepackage{amsmath,amsthm,amssymb}
\usepackage{array}
\usepackage{graphicx}
\usepackage{fancyhdr}
\usepackage{color}
\usepackage{extarrows}
\usepackage{subfigure}
\usepackage{enumerate}
\usepackage{multirow}
\usepackage{longtable}
\usepackage{setspace}
\usepackage{epsfig}
\usepackage[figuresright]{rotating}

\begin{document}

\title{On Uniqueness of ``SDE Decomposition'' in A-type Stochastic Integration}
\author{Ruoshi Yuan}
\affiliation{Key Laboratory of Systems Biomedicine, Ministry of Education, Shanghai Center for Systems Biomedicine,\\ Shanghai Jiao Tong University, Shanghai, 200240, China}
\author{Ying Tang}
\affiliation{Department of Physics and Astronomy, Shanghai Jiao Tong University, Shanghai 200240, China}
\author{Ping Ao}
\affiliation{Key Laboratory of Systems Biomedicine, Ministry of Education, Shanghai Center for Systems Biomedicine,\\ Shanghai Jiao Tong University, Shanghai, 200240, China}

\begin{abstract}

An innovative theoretical framework for stochastic dynamics based on a decomposition of a stochastic differential equation (SDE) has been developed with an evident advantage in connecting deterministic and stochastic dynamics, as well as useful applications in physics, engineering, chemistry and biology. It introduces the A-type stochastic integration for SDE beyond traditional Ito's or Stratonovich's interpretation. Serious question on its uniqueness was recently raised. We provide here both mathematical and physical demonstrations that the uniqueness is guaranteed. Such discussion leads to a better understanding on the robustness of the novel framework. We also discuss the limitation of a related approach of obtaining potential function from steady state distribution.   

\end{abstract}

\maketitle

Recently Zhou and Li (ZL) \cite{zhou2016construction} extensively discussed connections among three known ways of finding potential landscapes in generic nonequilibrium processes in biology, chemistry, engineering and physics described by stochastic differential equations. Some good questions were formulated, along with a few insightful results. They speculated that a set of differential equations proposed 10 years ago, referred to as ``SDE decomposition'' by ZL, would have generally no unique solution. In this comment we show that such speculation is not supported by either mathematical or physical reasoning. A few more points raised by ZL are further clarified.  

 We start by review the original definition \cite{ao2004,yuan2012beyond} of the  ``SDE decomposition'' into three components: a potential function $\phi(\mathbf{q})$ (a scalar function), a friction matrix $S(\mathbf{q})$ (symmetric and semi-positive definite) and a Lorentz-like force represented by a transverse matrix $A(\mathbf{q})$ (antisymmetric).  Matrices $S(\mathbf{q})$ and $A(\mathbf{q})$ are determined by the potential condition Eq.~\eqref{curl} and the generalized Einstein relation Eq.~\eqref{GER}:
\begin{align}
\label{curl}    &\nabla\times\left\{\left[ S(\mathbf{q})+A(\mathbf{q})\right]\mathbf{f}(\mathbf{q})\right\} = 0~, \tag{1a}\\
\label{GER} [S(\mathbf{q})&+A(\mathbf{q})]D(\mathbf{q})[S(\mathbf{q})-A(\mathbf{q})]=S(\mathbf{q})~.\tag{1b}
\end{align}
where $\mathbf{f}(\mathbf{q})$ is the deterministic drift velocity and $D(\mathbf{q})$ the diffusion matrix given by the SDE.
The $\nabla\times \mathbf{x}=\partial_{i}x_{j}-\partial_{j}x_{i}$ for arbitrary n-dimensional vector $\mathbf{x}$.
In principle, the $n^2$ unknowns in $[S(\mathbf{q})+A(\mathbf{q})]$ can be determined by solving the $n(n-1)/2$ partial differential equations in Eq.~\eqref{curl} \textit{under proper boundary conditions for matrices $S(\mathbf{q})$ and $A(\mathbf{q})$}, together with the $n(n+1)/2$ equations given
by Eq.~\eqref{GER} ($n^2$ unknowns and $n^2$ equations). Eq.~\eqref{curl} and Eq.~\eqref{GER} are the same as Eq.~(28a) and (28b) in \cite{zhou2016construction}. 

\setcounter{equation}{1}

Equation (1) may be transformed into a more standard but equivalent form. Notice that Eq.~\eqref{GER} implies a relation that $[S+A]=[D+Q]^{-1}$, where $Q$ is an antisymmetric matrix. We therefore rewrite Eq.~\eqref{curl} and Eq.~\eqref{GER} as
\begin{align}
\label{combined}
  & \nabla\times\left\{\left[ D(\mathbf{q}) + Q(\mathbf{q}) \right]^{-1} \mathbf{f}(\mathbf{q})\right\} = 0~
  \end{align}
where $n(n-1)/2$ partial differential equations determine the $n(n-1)/2$ unknowns in the antisymmetric matrix $Q$ with necessary boundary conditions for $Q$. One class of boundary conditions has been employed is that near fixed points every component of $Q$ is a smooth function of state variable \cite{ao2004,yuan2012beyond}. It can be assumed without lost of generality that both $D$ and $\mathbf{f}$ are smooth functions of state variable. Based on theory of partial differential equations, the existence and uniqueness of the solution is guaranteed at least locally \cite{pazy2012semigroups}. It should be pointed out that this general reasoning for uniqueness was also used by ZL for the solution of the Hamilton-Jacobi equation for the potential function \cite{zhou2016construction,yuan2012beyond}:
\begin{align}
\label{HJ}
\mathbf{f}(\mathbf{q})\cdot \nabla \phi (\mathbf{q})+\nabla \phi (\mathbf{q})\cdot D(\mathbf{q})\cdot \nabla \phi (\mathbf{q})=0~.
\end{align}

The existence and uniqueness of solution to Eq.~(1) has been explicitly and rigorously demonstrated for a class of important stochastic processes: linear stochastic processes including that of Ornstein-Uhlenbeck. The boundary condition of Eq.~(1) allows a gradient expansion scheme \cite{ao2004,yuan2012beyond} and for linear stochastic processes all higher orders are zero. If $\mathbf{f}(\mathbf{q}) = F \mathbf{q}$ and any two eigenvalues $\lambda_{i}$ and $\lambda_{j}$ of matrix $F$ satisfying $\lambda_{i}+\lambda_{j}\neq0$, the existence and uniqueness of $[S+A]$ ($S$ and $A$ are independent of state variable provided as boundary condition) in the whole state space are guaranteed by a theorem on Lyapunov equation \cite{ruo2014lyapunov,Haddad2008}. Note that the condition $\lambda_{i}+\lambda_{j}\neq0$ is not essential as already demonstrated \cite{Kwon2005} and explicit expression were obtained. Hence, the speculation on the general non-uniqueness by ZL is incorrect. 

The uniqueness of solution of Eq.~(1) may be suggested from physics side, too. The two fluctuation-dissipation relations define the diffusion matrix $D$ and friction matrix $S$, respectively. The transverse matrix $A$ can be related to an effective magnetic field. The physical meanings of $S$ and $A$ were apparently not used by ZL. They can be in principle measured independently. It is unlikely in a real physical process that two sets of results could be obtained. 
 
With the uniqueness of SDE decomposition settled generally, we turn to the toy model of a diffusion process on the circle $\mathcal{S}[0,~1]$ in ZL \cite{zhou2016construction}. We noted that it indeed reveals an additional feature largely overlooked, apparently even not noticed by ZL: The steady state distribution is generally not determined by the potential function in the form of Boltzmann-Gibbs distribution. There is no question to have Hamiltonian or potential function for dissipative dynamics in such situation \cite{ao1992influence}, and the steady state distribution can even be exactly evaluated \cite{chen1988return}. 
The mathematical reason of this mismatch is that such example brings the multi-connected state space into focus, where the solution to the potential condition, Eq.~\eqref{curl}, is sensitive to such topological constraint \cite{PhysRevE.92.062129}. In fact, the potential function can be formulated in different ways when the steady state distribution stays the same: For the toy model ZL has already shown that the decomposition (Eq.~(26) in ZL) straightforwardly holds with $\phi^{'}(x) = 0$, which leads to $\phi^{AO}(x)=constant$ (with $S=0$ by the generalized Einstein relation, and $D=1$, $A \equiv 0$) satisfying the global topological constraint. It can also be demonstrated that the potential function in the winding number representation, a washboard potential $\phi(x)= - x$ (with $S=1$ by the generalized Einstein relation), is also valid for the dynamics, revealing thermodynamical information such as nonequilibrium work when driven by an external control parameter \cite{PhysRevE.92.062129}. Note that washboard potentials exist in real physical systems \cite{chen1988return}, there is no reason to preclude them. They can be solutions to the Hamilton-Jacobi equation. Actually, two solutions of the Hamilton-Jacobi equation nicely correspond to the two of the generalized Einstein relation. Contrast to what ZL believed non-equivalence stated in their Table~IV,
this is an exactly demonstrable example of their asserted mathematical equivalence between the SDE decomposition theory and the Freidlin-Wentzell formulation.  

Having showed that ZL's non-uniqueness speculation is incorrect, several points raised in their paper deserve further discussion, clarification, and correction. 

\begin{itemize}
\item  
 First of all, we agree with the reasoning of ZL that the potential function should exist for nonequilibrium processes. It can be obtained by solving the Hamilton-Jacobi equation. This consensus may be important in that, while various construction of potential function are proposed recently \cite{ao2004,Wang2008, ao2013theory, cao2010probability, zhou2012quasi, yuan2012beyond, lu2014construction,qian2014zeroth}, there has been a lingering concern on its existence in mathematical community \cite{Strogatz2000}.    
\item  
 ZL noticed that the Hamilton-Jacobi equation plays a central role in the ``SDE decomposition'', similar to that in Freidlin-Wentzell formulation. We are pleased to notice such observation and wish to add that it was noticed by us, and the same Hamilton-Jacobi equation was written down, too \cite{yuan2012beyond}.
\item 
 ZL proved that the singularity on $D$ does not affect the decomposition framework.  We have also noticed and stressed this feature \cite{yuan2012beyond}, which is evident from the SDE decomposition but less obvious in other formulations. For example, a naive implementation of Freidlin-Wentzell formulation involved $1/D$ (c.f. Eq.~(12) of ZL), which apparently requires the non-singularity of $D$.  
\item 
 ZL was right in their observation that, though various proposals on potential landscape were made, no extensive discussion on their connections was performed. Nevertheless, we would remark that we had made an effort, particular in the connection between the A-type to that of Ito \cite{ao2007existence,shi2012}, and showed that mathematically they are equivalent. 
\item 
 Potential function is more generally applicable than the steady state distribution. In cases the steady state distribution does not exist, a potential function may still be obtainable. We had considered such case recently \cite{xu2014two}.  It is our observation that this important feature has not been generally appreciated so far in literature---ZL even attempted to show its equivalence to the Freidlin-Wentzell formulation. This feature, together with above discussion on the toy model, may lead to the conclusion that there is a serious limitation on its scope on the application of construction of potential function from steady state distribution.  
 \item
 One way to view the toy model on the ring is to consider it right at the limit cycle in a limit cycle dynamics, as stressed by ZL. This explicitly implies that there is no dissipation right at the limit cycle, $S=0$, and that the dynamics is dominated by non-gradient term, or the transverse matrix in the SDE decomposition. Not noted by ZL, such result was also exactly obtained within the SDE decomposition \cite{Zhu2006,yuan2013exploring}.  
 
\item 
 ZL erroneously referred the fluctuation-dissipation theorem \cite{yuan2012beyond} $\tilde{\sigma}(x)\tilde{\sigma}(x)^{t}=2\varepsilon S(x)$ as the generalized Einstein relation \eqref{GER}. In case with detailed balance, $A = 0$, let $\epsilon=k_BT$, if the friction $\gamma$ is a constant, then $S=\gamma/k_BT$, Eq.~\eqref{GER} reduces to $SD=\gamma D/k_BT = 1$, namely, the product of the friction and diffusion coefficients is a constant $\gamma D=k_BT$, discovered by Einstein \cite{Einstein1905,Kubo1966} one hundred years ago.

\item 
 ZL remarked that ``the general mathematical expression for OM function in high dimensional cases has been studied in Refs.~39 and 40, which include the result in Ref.~38 as a special case.'' According to our knowledge, however, Refs.~39 and 40 in \cite{zhou2016construction} provide just an action function under Stratonovich's interpretation without any explicit action function for the general $\alpha$-type interpretation even for the one dimensional case. 
In contrast, we find there are multiple consistent explicit action functions under the $\alpha$-type interpretation correspond to different integration rules \cite{tang2014summing}, clarifying confusions in the field.
\item 
 ZL stated that for SDE decomposition ``there is no explicit stochastic integral interpretation of it in higher dimensions''. It is an incorrect assertion. The zero mass limit justification \cite{Yin2006} of SDE decomposition is in fact an explicit realization of stochastic integration: first the usual, for example, Ito stochastic integration and then the zero mass limit. It is of course not the usual stochastic integration in a standard textbook, as we have already recognized as beyond Ito vs Stratonovich \cite{yuan2012beyond}. It is possible that a more conventional form may be found.   
\item 
  ZL asked the important question of how to generalize what has been obtained for continuous processes to discrete jump processes and speculated that the ``SDE decomposition'' theory would be difficult to do that. We had in fact showed the generalization is possible \cite{ping2013dynamical}. Such generalization is a direct extension of the ``SDE decomposition''.    

\end{itemize}

We also provide a detailed discussion on ZL's demonstration in the appendix. We note: 
\begin{itemize}
\item[a)] The starting point of ZL is not the original definition of the ``SDE decomposition'' by Eq.~(1).  ZL's protocol is not equivalent to the original definition: necessary boundary conditions for $Q$ are missing. This can naturally explain the freedom they found in their Theorem 2. 
\item[b)] The freedom reported by ZL in choosing dynamical matrices corresponds to cancelled-out forces shown in Fig.~(1) that has a similar role as the gauge freedom in electromagnetics or simply a reference point for a potential function which needed to be specified by boundary conditions.
\item[c)] The steady state distribution of A-type FPE is unique even without boundary conditions for $Q$, the freedom observed in ZL's Theorem 2 actually show the robustness of the A-type framework.
\end{itemize}
To summarize, while ZL reported many interesting observations, they have not cited work careful enough.
Their non-uniqueness speculation is not only un-rigorous, it is also incorrect.   
\bibliography{Yrs1}
\newpage
\appendix
\section{On Zhou and Li's demonstration}
The starting point of Zhou and Li in \cite{zhou2016construction} is not the original definition of the ``SDE decomposition'' by Eq.~(1). They instead used Hamilton-Jacobi equation to solve the potential function $\phi(\mathbf{q})$, an equation explicitly eliminating the antisymmetric matrix $Q(\mathbf{q})$. And then, they try to use $\phi$ to reconstruct dynamical matrix $Q$ and find that there are multiple choices of $Q$ in their Theorem 2.
We note that this protocol is not equivalent to the original definition: necessary boundary conditions for $Q$ are missing. In fact, a rough estimation of the freedom find in their Theorem 2 is straightforward: there are $(n-1)n/2$ boundary conditions missing for solving $Q$ in \eqref{combined}, but specified $n-1$ boundary conditions in solving $\phi$ using Hamilton-Jacobi equation, the number of  undetermined freedom is $(n-1)n/2-(n-1)=(n-1)(n-2)/2$ corresponding to their conclusion in Theorem 2.

From a physical viewpoint, the problem of missing boundary conditions becomes more evident. Let us consider a symmetric problem
\begin{eqnarray}
  &\left[ S(\mathbf{q})+A(\mathbf{q})\right]\mathbf{f}(\mathbf{q})=\left[ S'(\mathbf{q})+A'(\mathbf{q})\right]\mathbf{f}(\mathbf{q})=-\nabla \phi(\mathbf{q})~,&\label{e1}\\
  &S'(\mathbf{q})=S(\mathbf{q})+\Delta S(\mathbf{q})~,\quad A'(\mathbf{q})=A(\mathbf{q})+\Delta A(\mathbf{q})~,\quad\Delta S(\mathbf{q})\mathbf{f}(\mathbf{q})=-\Delta A(\mathbf{q})\mathbf{f}(\mathbf{q})&\label{ee}
\end{eqnarray}
by given $\phi(\mathbf{q})$, there are obvious multiple choices of $S$ and $A$, such as $S'$ and $A'$ satisfying \eqref{ee}. The physical meaning here is straightforward: the frictional force $S(\mathbf{q})\mathbf{f}(\mathbf{q})$, the Lorentz force $A(\mathbf{q})\mathbf{f}(\mathbf{q})$ and the electrostatic potential $\phi(\mathbf{q})$ can be tuned independently in an experiment. Given solely the electrostatic potential $\phi(\mathbf{q})$, one cannot determine the experimental system--there can exist an arbitrary but cancelled-out force between the frictional force and the Lorentz force $\Delta S(\mathbf{q})\mathbf{f}(\mathbf{q})=-\Delta A(\mathbf{q})\mathbf{f}(\mathbf{q})$, as shown in Fig.~1. To specify the real situation, boundary conditions for the friction and the magnetic field should be provided as in Eq.~\eqref{curl} and Eq.~\eqref{GER}. The case for $Q$ is similar by the symmetry between the Eq.~\eqref{e1} and Eq.~\eqref{b1} (Eq.~(B1) in \cite{zhou2016construction}).

\begin{figure}
\begin{centering}
\includegraphics[width=0.5\textwidth]{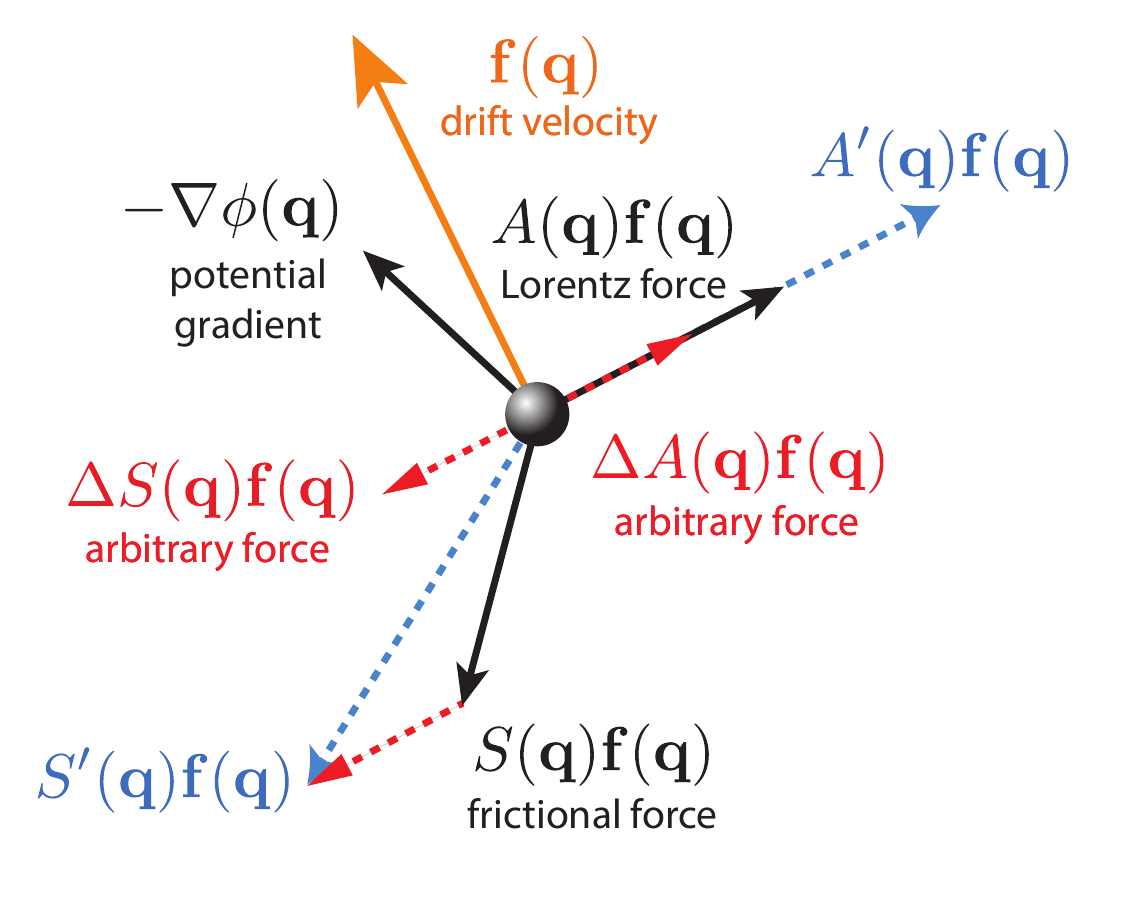}
\caption{Without necessary boundary conditions, there are multiple physical realizations of Eq.~\eqref{e1}. 
The extra freedom corresponds to arbitrary cancelled-out forces perpendicular to the drift velocity.
These realizations do not lead to any physically observable consequence.}
\par\end{centering}
\end{figure}

Furthermore, regardless of the multiple choices of $Q(\mathbf{q})$ caused by the missing boundary conditions, the steady state distribution for A-type Fokker-Planck equation is still uniquely determined. The A-type Fokker-Planck equation
\begin{align}
\label{FPE}
  \partial_t \rho(\mathbf{q},t)=\nabla\cdot\left[D(\mathbf{q})+Q(\mathbf{q})\right]\cdot[\epsilon\nabla+\nabla\phi(\mathbf{q})]
    \rho(\mathbf{q},t)~,
\end{align}
has a steady state solution (if normalizable)
\begin{align}
\label{BG}
   \rho_s(\mathbf{q}) = \frac{1}{Z(\epsilon)}\exp\left(-\frac{\phi(\mathbf{q})}{\epsilon}\right)~,
\end{align}

For the multiple solutions $Q$ and $Q'$ of (B1) in \cite{zhou2016construction},
\begin{eqnarray}
   [D(\mathbf{q})+Q(\mathbf{q})]\nabla \phi(\mathbf{q})= [D(\mathbf{q})+Q'(\mathbf{q})]\nabla \phi(\mathbf{q})=-\mathbf{f}(\mathbf{q})\label{b1} \\
       Q'(\mathbf{q})=Q(\mathbf{q})+\Delta Q(\mathbf{q})~,\quad\Delta Q(\mathbf{q})\nabla \phi(\mathbf{q})=0~.\label{qq}
 \end{eqnarray}
 Since $\nabla\rho_s(\mathbf{q})\propto\nabla\phi(\mathbf{q})\cdot\rho_s(\mathbf{q})$, Eq.~\eqref{BG} is still a steady state solution of the A-type Fokker-Planck equation \eqref{FPE} with arbitrary $Q'$. Since obtaining steady state distribution is enough in many applications, e.g., calculating spontaneous transition rates between stable states, A-type framework is able to generate the steady state distribution without specifying boundary conditions for the matrices.

\end{document}